\documentclass{iau} 
\usepackage{graphicx}

\title[Cosmic evolution of higher radiative luminosity AGN] 
{Cosmic evolution of AGN with moderate-to-high radiative luminosity in the COSMOS field
}

\author[Lana Ceraj et al.]   
{L. Ceraj$^1$ \and V. {Smol\v{c}i\'{c}}$^1$ \and I. Delvecchio$^1$ \and J. Delhaize$^1$ and M. Novak$^1$
}

\affiliation{$^1$University of Zagreb, Physics Department, Bijeni\v{c}ka cesta 32, 10002 Zagreb, Croatia}

\pubyear{2018}
\volume{333}  
\jname{Peering towards Cosmic Dawn}
\editors{Vibor Jeli\'c \& Thijs van der Hulst, eds.}
\begin{document}

\maketitle

\begin{abstract}
We study the moderate-to-high radiative luminosity active galactic nuclei (HLAGN) within the VLA-COSMOS 3 GHz Large Project. The survey covers 2.6 square degrees centered on the COSMOS field with a 1$\sigma$ sensitivity of 2.3 $\mathrm{\mu Jy}$/beam across the field. This provides the simultaneously largest and deepest radio continuum survey available to date with exquisite multi-wavelength coverage. The survey yields 10,830 radio sources with signal-to-noise ratios $\geq$5. A subsample of 1,604 HLAGN is analyzed here. These were selected via a combination of X-ray luminosity and mid-infrared colors. We derive luminosity functions for these AGN and constrain their cosmic evolution out to a redshift of $z\sim6$, for the first time decomposing the star formation and AGN contributions to the radio continuum emission in the AGN. We study the evolution of number density and luminosity density finding a peak at $z\sim1.5$ followed by a decrease out to a redshift $z\sim6$.
\keywords{Surveys; galaxies: general, active; separation:
	general; radio continuum: galaxies; X-rays: galaxies: AGN}
\end{abstract}

\firstsection 

\section{Introduction}

It is now widely accepted that the evolution of host galaxies is related to the evolution of the active galactic nuclei (AGN) within them. Therefore, understanding the evolution of galaxies requires understanding how their components change through cosmic time. 

Multi-wavelength studies show a dichotomy in the properties of galaxies containing an AGN. This AGN dichotomy is thought to reflect a difference in the efficinecy of accretion onto the central supermassive black hole (SMBH; e.g. see review by \cite[Heckman \& Best 2014]{BH14}). Optical emission lines are one of the spectral signatures of radiatively efficient accretion due to the inflow of the cold gas onto the SMBH. Objects containing this type of signature are referred to as high-excitation radio galaxies (HERGs) or radiative-mode AGN (see eg. Tab. 6 in Smol\v{c}i\'{c} 2016).
In the case of inefficient accretion fueled by the hot gas, optical emission lines are often very weak. However, inefficient accretion can produce powerful jets which can outshine the host galaxy in radio. These types of objects are called jet-mode AGN or low-excitation radio galaxies (LERGs). Radio jets may occur also in radiative-mode AGN, but their radiation is not as strong as in the jet-mode AGN. 

In this work we study the cosmic evolution of a population of X-ray and mid-infrared (MIR) selected AGN out to a redshift $z\sim6$ in the COSMOS field, detected within the VLA-COSMOS $3$ GHz Large Project data (Smol\v{c}i\'{c} et al. 2017a). This selection aims to trace the high redshift analogs of HERGs. Previous studies conducted on radio selected samples of X-ray and mid-IR AGN suggest that AGN identified through these criteria are hosted mostly by star-forming galaxies (SFGs) in which the bulk of radio emission originates from the host galaxy rather than SMBH activity (eg. Padovani et al. 2015). For instance, Delvecchio et al. (2017) report that $\sim$30$\%$ of the moderate-to-high radiative luminosity AGN (HLAGN) display significant radio excess due to non-thermal AGN emission, while the rest are dominated by star formation in radio.

To constrain the evolution of the AGN-related emission, one needs to decompose the total radio emission into the components that arise from AGN activity and star forming processes.
In this work we develop a statistical method for decomposing the radio luminosity onto the star forming and AGN contributions, which is crucial to trace the evolution of radio emission associated only with the AGN. 
We construct the AGN-related radio luminosity functions, which we further use to constrain the pure density (luminosity) evolution of our X-ray and MIR selected AGN. 

\section{Data}

We are using radio data from the VLA-COSMOS 3 GHz Large Project (Smol\v{c}i\'{c} et al. 2017a). The 3 GHz data from the observations of the 2.6 $\mathrm{deg^2}$ sky area centered on the COSMOS field yielded the detection of 10,830 sources with signal-to-noise ratio $\geq$5. 

By cross-matching the radio sample with the catalog of multiwavelength data (COSMOS2015; Laigle et al. 2016), a sample of 7,729 radio sources have been assigned optical/ near-IR counterparts (see Smol\v{c}i\'{c} et al. 2017b). These sources have further been classified based on their X-ray luminosity, MIR and optical colors into three subcategories: HLAGN (moderate-to-high radiative luminosity AGN), MLAGN (low-to-moderate radiative luminosity AGN) and SFGs. For more information on the classification scheme and the analysis of the properties of these sources, we refer the reader to Smol\v{c}i\'{c} et al. (2017b) and Delvecchio et al. (2017). 

In our work we are using the HLAGN sample of 1,604 radio sources. 

\section{Decomposition and radio luminosity functions of HLAGN}

We expect radio sources classified based on their X-ray and MIR emission to be hosted by galaxies in which significant amount of the radio emission originates from star formation processes, while only a fraction originates in AGN (Padovani et al. 2015, Delvecchio et al. 2017). To properly constrain the evolution of AGN in these sources, the radio emission has to be disentangled into that originating from star-forming processes and AGN activity. 

We decompose the total 1.4 GHz luminosity using the infrared-radio correlation (IRRC) derived by Delhaize et al. (2017). They examined the cosmic evolution of the IRRC using a sample of $9,575$ star-forming galaxies in the COSMOS field. These objects display no evidence of AGN presence (Smol\v{c}i\'{c} et al. 2017b). They find a redshift dependent infrared-to-1.4 GHz radio luminosity ratio ($q_{TIR}$) with a spread 1$\sigma$=0.35 dex, which slightly decreases towards higher redshifts. 

All sources within $\pm 2\sigma$ from the IRRC are expected to have radio emission dominated by star formation processes. We define AGN fractions by quantifying an excess of radio emission for the sources below $q_{TIR}-2\sigma$, performing Monte Carlo simulations to account for the uncertainties of IRRC and luminosities used in the calculations. We find that HLAGN span a wide range of AGN fractions, ranging from galaxies dominated by star formation processes to those in which the main contribution to the radio emission is due to AGN activity.

\begin{figure}[h]
	\includegraphics[width=\linewidth]{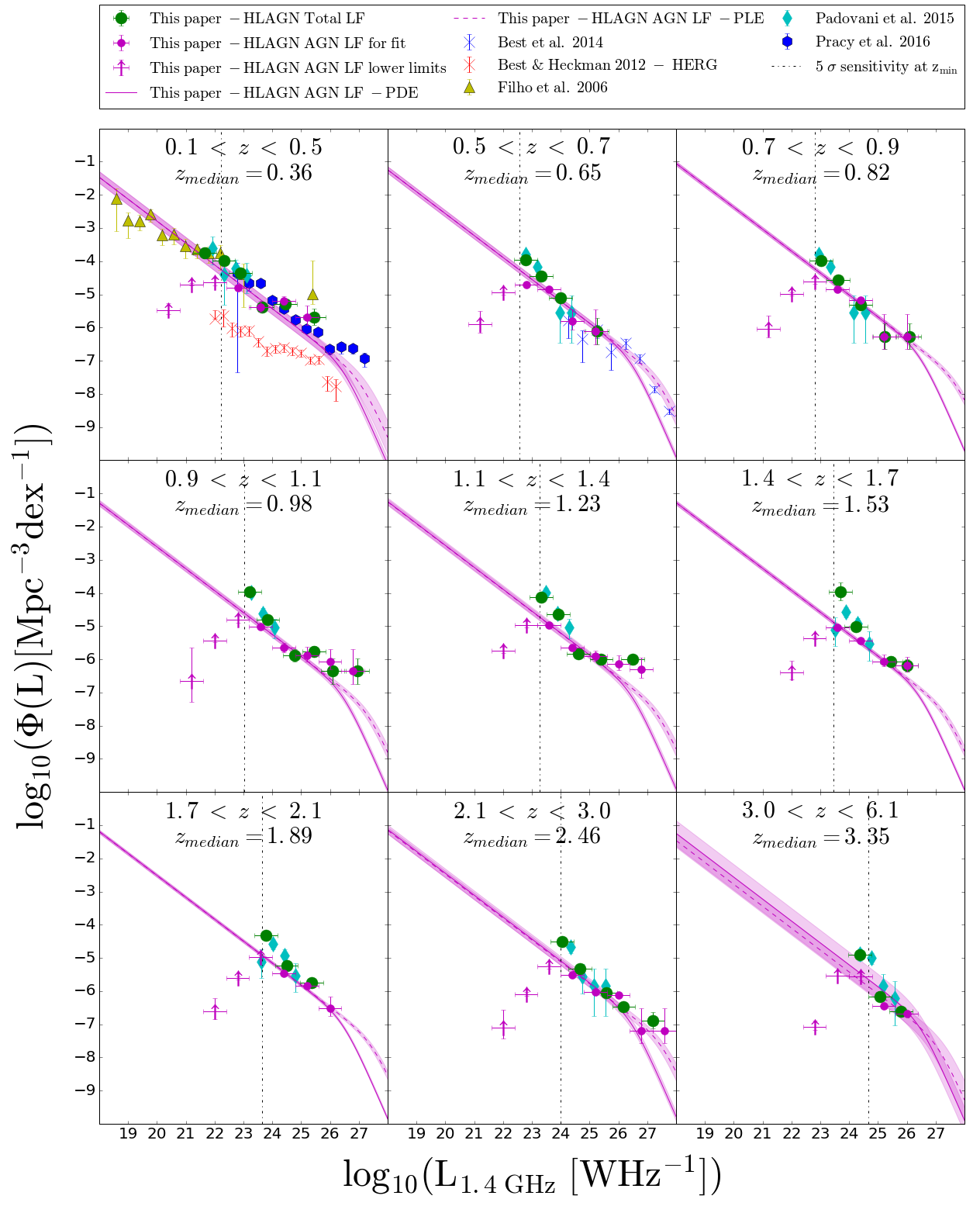}
	\caption{The AGN luminosity functions of the HLAGN sample. Magenta dots and lines show the data and analytical fit to the data, respectively. Green dots show the total HLAGN luminosity functions.}
\end{figure}

We further derive the AGN luminosities by scaling the total 1.4 GHz luminosities by the AGN fractions. We separate the HLAGN sample into 9 redshift bins and calculate the maximum observable volume in which each of the galaxies in our sample could be detected using the standard $\mathrm{V_{max}}$ method (Schmidt 1968). We also mitigate completeness issues and take into account the geometry of the survey (the effective unmasked area of the COSMOS optical-NIR survey is 1.77 $\mathrm{deg^2}$). Using the maximum observable volumes and AGN luminosities, we construct the 1.4 GHz AGN luminosity functions. We perform Monte Carlo simulations to estimate the overall uncertainties in the derivation of AGN luminosities. To illustrate how the luminosity decomposition affects the shape of luminosity functions, we also calculate the total 1.4 GHz luminosity functions following a procedure consistent with that presented in Novak et al. (2017) and Smol\v{c}i\'{c} et al. (2017c).

Our AGN luminosity functions in the first redshift bin (0.1$<z<$0.5) agree well with the local HERG luminosity functions derived by Pracy et al. (2016). They used a sample of radio galaxies identified from matching FIRST sources with SDSS images. Based on their optical spectra, sources were classified into LERG and HERG classes (see Ching et al. 2017). By statistically decomposing their HERG luminosity functions, we find that they are dominated by AGN emission and are in excellent agreement with our estimates over the same range of 1.4 GHz luminosities.

To constrain the evolution of HLAGN from redshift $z=0.1$ to $z\sim6$ we fit the analytic expression given in Pracy et al. (2016) for the HERG luminosity function. We test the pure density and pure luminosity evolution models, in both cases finding a decline of the evolution parameters towards higher redshifts. We further use these evolution parameters to calculate number and luminosity densities of HLAGN. Both curves increase from high redshifts up to a peak at $z\sim1.5$, and display a decline towards $z=0$.

Using the scaling relation between monochromatic luminosity and kinetic luminosity by Willott et al. (1999), we calculate the kinetic luminosity density. Assuming an uncertainty factor $f_W=15$, we find a good agreement between predictions from semi-analytical models of radio AGN feedback (Croton et al. 2016) and our derivation.

\section{Summary and conclusions}

In this work we used a sample of 1,604 X-ray and mid-IR selected AGN with moderate-to-high radiative luminosities (HLAGN) from the VLA-COSMOS 3 GHz Large Project. By quantifying an excess of radio emission from the IRRC derived by Delhaize et al. (2017), we decompose the total 1.4 GHz radio luminosity onto the luminosity arising from an AGN and star formation processes. Using the AGN luminosities, we construct AGN luminosity functions in 9 redshift bins up to $z\sim6$ and trace its evolution in the case of pure density (luminosity) models. We further calculate the number (luminosity) density evolution of HLAGN, finding a peak at redshift $z\sim1.5$.

\end{document}